\documentclass[aps,prd,twocolumn,aps,floatfix,showpacs,showkeys,superscriptaddress,nofootinbib]{revtex4-1}

\usepackage{amssymb}
\usepackage{amsbsy}
\usepackage{amsthm}
\usepackage{amsmath}
\usepackage{breqn}
\usepackage{bm}
\usepackage{braket}
\usepackage{color}
\usepackage{dcolumn}
\usepackage{epsfig}
\usepackage[T1]{fontenc}
\usepackage{graphicx}
\usepackage[colorlinks]{hyperref}
\usepackage[latin9]{inputenc}
\usepackage{mathtools}
\usepackage{placeins}
\usepackage{subcaption}
\usepackage{xcolor}
\usepackage{cleveref}
\begin{document}
	
\title{Locality \& Entanglement in Table-Top Testing of the Quantum Nature of Linearized Gravity}

\newcommand{\affone}{Department of Physics and Astronomy, University College London, Gower Street, WC1E 6BT London, United Kingdom.}
\newcommand{\afftwo}{Van Swinderen Institute, University of Groningen, 9747 AG Groningen, The Netherlands.}
\newcommand{\affthree}{Department of Physics, University of Warwick, Gibbet Hill Road, Coventry CV4 7AL, United Kingdom.}
	
\author{Ryan J. Marshman}
\affiliation{\affone}
	
\author{Anupam Mazumdar}
\affiliation{\afftwo}

\author{Sougato Bose}
\affiliation{\affone}
	
\date{\today}

\begin{abstract}
This paper points out the importance of the assumption of locality of physical interactions, and the concomitant necessity of propagation of an entity (in this case, off-shell quanta -- virtual gravitons) between two non-relativistic test masses in unveiling the quantum nature of linearized gravity through a laboratory experiment. 
At the outset, we will argue that observing the quantum nature of a system is not limited to evidencing $O\left(\hbar\right)$ corrections to a classical theory: it instead hinges upon verifying tasks that a classical system cannot accomplish. We explain the background concepts needed from quantum field theory and quantum information theory to fully appreciate the previously proposed table-top experiments: namely forces arising through the exchange of virtual (off-shell) quanta, as well as Local Operations and Classical Communication (LOCC) and entanglement witnesses. We clarify the key assumption inherent in our evidencing experiment, namely the locality of physical interactions, which is a generic feature of interacting systems of quantum fields around us, and naturally incorporates micro-causality in the description of our experiment. 
We also present the types of states the matter field must inhabit, putting the experiment on firm relativistic quantum field theoretic grounds. At the end we use a non-local theory of gravity to illustrate how our mechanism may still be used to detect the qualitatively quantum nature of a force when the scale of non-locality is finite. We find that the scale of non-locality, including the entanglement entropy production in local/ non-local gravity, may be revealed from the results of our experiment.

\end{abstract}

\maketitle

\section{Introduction}

Recently, two papers~\cite{Bose:2017nin, Marletto2017} have discussed 
the possibility of detecting quantum behavior of a linearized gravitational field in a table top experiment. 
The proposal crucially relies on local quantum interactions between matter and the gravitational field leading to the generation of entanglement between the two non-relativistic test masses, each initially prepared in a superposition of distinct spatial states. This entanglement is a proof of the quantum-ness of the mediating gravitational field, and can be witnessed by measuring the correlations between individual spins which have been embedded in the test masses \cite{Bose:2017nin}. The witness can be measured in a few runs of the experiment if the entanglement generating phase 
due to this gravitational potential between the two superposed quantum systems is roughly of order one. While the proposed experiment had been couched in terms of Stern-Gerlach interferometry~\cite{Folman2018,Wan2016} which enabled its formulation in terms of a spin entanglement witness, it is possible that other settings in which macroscopic superpositions are generated will work just as well~\cite{marinkovic2018optomechanical, ahn2018optically, kaltenbaek2016macroscopic, arndt2014testing,miao2019quantum,krisnanda2019observable,bykov2019direct,qvarfort2018mesoscopic}.

For the conclusion about the quantum nature of gravity to follow from the aforementioned entanglement, it is very important that ``something'' is exchanged between the test masses when they interact mutually through their Newtonian interaction. This point is often unclear when the proposed entanglement generation experiment is presented in terms of a direct Newtonian interaction between the test masses resulting in appropriate phase evolutions in their states which entangle the masses.  In fact, that approach is adopted purely for convenience and we highlight here that there is a very well defined quantum mechanism for the Newtonian interaction where the entity which acts as a mediator of the interaction is an off-shell (virtual) graviton. This is exchanged between the test masses, and through a tree-level diagram leads to the Newtonian interaction.   Fundamentally, according to quantum field theory, forces between two sources (say two static charges) can be understood from the exchange of virtual particles between them -- photons, $W^{\pm}, Z$ bosons and gluons -- which are uncontroversially (by definition) quantum mechanical~\cite{srednicki2007quantum}. Similarly, in the low curvature regime, gravity can be regarded as perturbations on a background, and these perturbations can be regarded as a field. Within this setting, the Newtonian interaction between two masses can be considered as originating from the exchange of virtual gravitons~\cite{feynman2018gravity}, which puts gravity, at least in this regime, in exactly the same quantum footing as the other known fundamental forces of nature. 

However, the mere theoretical existence of a quantum mechanism for the origin of the Newtonian force does not prove that it is indeed that quantum mechanism that nature has decided to adopt -- only experiments can do that. As far as current experimental evidences are concerned, it could equally well be a classical field generated by a source mass which affects a probe mass placed in that field -- indeed there are several proposed classical and semiclassical mechanisms to generate a force with the same features as the Newtonian force~\cite{moller1963theories, rosenfeld1963quantization, penrose1996gravity, diosi1987universal, kafri2014classical, kafri2015bounds, carney2018tabletop, tilloy2016sourcing,jacobson1995thermodynamics,oppenheim2018post}. How do we know whether any of these other mechanisms are adopted by nature or whether it is indeed the exchange of quantum off-shell gravitons? 

Detecting the quantum nature of an entity has historically been through radical ``qualitative" departures (such as the photoelectric effect detecting energy quantization) or through ``quantitative" $O(\hbar)$ quantum corrections to energies and interaction potentials. However, the above strategies hardly seem to be adaptable readily to the case of a laboratory test for the quantum nature of gravity. Its on-shell quantum wavepackets, the gravitons (say, of a gravitational wave), carry too little energy, while any $O(\hbar)$ quantum modifications of the Newtonian potential are too small to witness for currently available systems. Thus the question arises: could one cleverly design a laboratory experiment to reveal 
an underlying quantum mechanism of the Newtonian gravitational force itself.  Unfortunately, this underlying quantum character is completely {\em hidden} in the subset of experiments done so far: these look solely at the classical effects of the force field. Examples of this included the displacement of an object in a Newtonian potential or the phase development of a wavefunction of a quantum object in that classical field \cite{Colella}. Furthermore, previous suggestions regarding observation of gravitational effects cannot unambiguously falsify quantum gravity~\cite{ashoorioon2014implications}.

Thus, the recent papers have had to propose an indirect strategy \cite{Bose:2017nin, Marletto2017}. If an agent entangles two quantum entities, the agent must be performing quantum communication between them, i.e., it must itself be a quantum entity. Through this idea, the generation of entanglement between two masses is used to witness the quantum nature of the agent acting between them. Note that here we are testing a quantum feature of gravity in a similar spirit to a Bell-inequality test on quantum systems ~\cite{hensen2015loophole}, which is an effect that {\em does not go away} when $\hbar \rightarrow 0$, as was shown a long time ago using two entangled large spins \cite{Peres, Gisin-Peres}, although it might become more challenging to detect. Other similar quantum effects that survive as $\hbar \rightarrow 0$ have recently been proposed in the context of the violation of macro-realism  by large spins and large masses \cite{Home-Mal, Bose-Mal}. Similarly the effect we suggest  is a quantum effect that remains in the $\hbar \rightarrow 0$ limit, and while it is difficult to detect, we have suggested, in Ref.\cite{Bose:2017nin}, a domain in which it is feasible to be observed. Several viewpoints have been presented regarding the interpretation and applications of this idea~\cite{Bose:2017nin} (supplementary material), \cite{Marletto2017,carney2018tabletop, anastopoulos2018comment, hall2018two, marletto2018can, superpositionofgeometries2018, christodoulou2018possibility, belenchia2019information,giampaolo2018entanglement}. There have also been noise analysis~\cite{nguyen2019entanglement}, as well as related independent suggestions~\cite{carlesso2019testing,al2018optomechanical} and paradox resolutions~\cite{mari2016experiments, belenchia2018quantum} which point towards the necessity of gravity to be quantum in nature.

In this paper, we seek to clarify the crucial {\em assumptions} underlying the claim that the witnessing of entanglement in the laboratory demonstrates the quantum nature of gravity. Moreover, we will show that it all works consistently within  a quantum field theory context using a fully relativistically covariant formalism for the propagator. This also naturally clarifies how relativistic causality can be respected in the treatment of the above experiments. We start by laying down {\em all} our assumptions, the most important being the locality of physical interactions, in the above evidencing of quantum-ness. We clarify the manner in which the gravitational field would entangle the spins via the energy momentum tensor of the non relativistic mesoscopic superpositions. We will further clarify the necessity of the interaction to be through a quantum entity to allow such entanglement to form, by clarifying why local operations and classical communications (LOCC) cannot entangle the masses in our scenario. Specifically, as the term ``communication'' may sound somewhat cryptic to the physicist who thinks about interactions between fields, we show the impossibility of a classical gravitational field to create entanglement. The notion of classical field here is kept very general and automatically includes situations such as semi classical gravity (quantum matter sourcing a classical gravitational field) as well as were the matter is not strictly quantum mechanical in the usual sense - i.e., it has stochastic evolutions beyond standard quantum mechanics (e.g., when they are subject to fundamental collapse models) so that the gravitational field generated is also stochastic. As far as the experimental aspects are concerned, we emphasize why we seek the simplest statistical procedure to witness the entanglement rather than trying to estimate an entanglement measure. Given the fundamentally quantum field theoretic nature of all systems, one should also treat the test masses as described by quantum fields. In this context we present the type of states the matter field must be assumed to inhabit for a simple ``bipartite'' witnessing of the entanglement. Finally, adopting the example of a ``non-local'' theory of gravity, where there is a valid quantum propagator, we  provide an example where our method can still be used to witness the underlying quantum nature of the field even though the theory is fundamentally non-local at some scale. In fact, this example illustrates that as long as the length scale of non-locality is ``finite'', our mechanism is a valid approach as there is the need for entities to propagate from point to point to convey an interaction. Interestingly enough, our experiment can also be used to {\em reveal} the length scale of non-locality, if present.


\section[Assumptions]{Underlying Assumptions \label{sec:Assumptions}}
To begin with, it is worth highlighting the key assumption underlying the inference of the quantum nature of gravity from our tabletop experiment on gravitationally mediated entanglement
\begin{itemize}
\item{\bf Locality of physical interactions:} One of the pillars of quantum field theory is the assumption of locality. All the interactions are assumed to be local at both classical and at a quantum level. Locality also ensures micro-causality~\footnote{Specifically, the field operators for two masses $\hat{\phi}_{1}\left(\mathbf{x}_a\right)$ and $\hat{\phi}_{2}\left(\mathbf{x}_b\right)$, where $\mathbf{x}_i$ are the four vectors for masses, minimally coupled through the gravitational field can be considered. When the two masses are space-like separated
$\Delta s^2\left(\mathbf{x}_a -\mathbf{x}_b\right) > 0$
and
$\left[\hat{\phi}_{1}\left(\mathbf{x}_a\right),\hat{\phi}_{2}\left(\mathbf{x}_b\right)\right]=0$
and as such we have no faster than light signalling. Now of course when 
$\Delta s^2\left(\mathbf{x}_a -\mathbf{x}_b\right) < 0$
and
$\left[\hat{\phi}_{1}\left(\mathbf{x}_a\right),\hat{\phi}_{2}\left(\mathbf{x}_b\right)\right]\ne0$
and so all causal relationships will behave as expected. \label{footnote:causality}}.
In the context of gravity, the local interaction is given by:
\begin{equation}
\kappa^2 h_{\mu\nu}(\vec{r},t)T^{\mu\nu}(\vec{r},t) \label{eq:interaction}
\end{equation} 
where $\kappa^2=(8\pi G)^{-1}$, $G=\hbar/M_p^2$ is Newton's constant, $M_p\sim 10^{19}$GeV, $\mu,\nu=0,1,2,3$ and we are working with signature $(-,+,+,+)$. The energy momentum tensor of matter is given by $T_{\mu\nu}$. The metric perturbation around Minkowski background is
\begin{equation}
g_{\mu\nu}=\eta_{\mu\nu}+\kappa h_{\mu\nu}\,,
\end{equation}
where $\eta_{\mu\nu}$ is the Minkowski metric, and $|\kappa h_{\mu\nu}|\ll 1$, in order to maintain the linearity.
{\em A priori} $h_{\mu\nu}$ need not be quantum at all. Though the matter part of the energy momentum tensor could be a quantum entity.  

The concept of locality is also an important criteria from the perspective of quantum information and quantum entanglement. In particular, under LOCC, two particles exchanging only classical energy momentum will not lead to enhancement in entanglement. Note that while LOCC is used as a principle to define mixed state entanglement~\cite{horodecki1998mixed, bennett1996mixed}, it can be easily {\em proved} when we start from an unentangled state of two objects as in the case of the experiments described in Refs.\cite{Bose:2017nin,Marletto2017}. In fact in these experiments, the two applications of locality, i.e., in defining local quantum field theories and in prohibiting entanglement generation at a distance without quantum communication, are brought together. It is very important to note that the locality is not proven through our experiment -- that is not its purpose -- locality is assumed from our knowledge of physical interactions in the observed regimes. It is the quantum aspect which we prove after assuming the locality.

Of course, as opposed to a  local quantum  theory, non-local field theories have also been developed since the days of Yukawa~\cite{Yukawa:1950eq}, and Pais and Uhlenbeck~\cite{Pais:1950za}. There has been a recent resurgent in understanding them as well in the context of field theory~\cite{Efimov:1967pjn,Tomboulis:1997gg,Biswas:2005qr,Biswas:2011ar}, and in quantum mechanics~\cite{Buoninfante:2017kgj,Boos:2018kir,Buoninfante:2019teo}. Note these are however {\em not} ``action at a distance'' theories. One of the features of a non-local theory is that it does not have a point support~\cite{Siegel:2003vt,Biswas:2005qr,Buoninfante:2018rlq}, therefore it is very helpful towards ameliorating some of the singularities in nature, such as point singularity due to gravitational $1/r$ potential~\footnote{Infinite derivatives acting on delta Dirac source does not have a point support. Let us consider a one-dimensional problem,
\begin{equation}
e^{\alpha\nabla^2_x}\delta(x)=\frac{1}{\sqrt{2\pi} }\int dk e^{-\alpha k^2} e^{ik\dot x}=\frac{1}{\sqrt{2\alpha}}e^{-x^2/4\alpha}\,,
\end{equation}
Note that the left hand side is a non-local operator acting on a delta Dirac source, with a scale of non-locality given by $\alpha^{-1}$. The result is a Gaussian distribution. In a very similar fashion one can also resolve the singularity present in a rotating metric in general relativity~\cite{Buoninfante:2018xif}, and the singularity due to a charged electron~\cite{Buoninfante:2018stt}.
},
\footnote{The non-local theories arise in many contexts in quantum gravity, in string theory, the notion of point objects are replaced by strings and branes~\cite{Polchinski:1998rr}, dynamical triangulation~\cite{Ambjorn:2012jv}, loop quantum gravity~\cite{Rovelli:1994ge}, and casual set approach~\cite{Surya:2019ndm} exploits Wilson operators which are inherently non-local. The string field theory introduces non-locality at the string scale, for a review~\cite{deLacroix:2017lif}, and infinite derivative  ghost free theory of gravity (IDG)~\cite{Biswas:2011ar}, which does not introduce any instability around a given background, is motivated from string field theory~\cite{Witten:1985cc,Freund:1987ck,Freund:1987kt}. Especially, in string field theory and in IDG the non-locality appears only at the level of interactions. Note that loss of locality will also give rise to violation of micro-causality. However, it has been shown that for a specific class of non-local theories we are interested in here,  the violation of causality is limited to the scale of non-locality~\cite{Efimov:1967pjn,Tomboulis:1997gg,Tomboulis:2015gfa,Buoninfante:2018mre}. }.

In this paper, as an alternative to local gravitational interaction, we will also study a non-local theory of gravity~\cite{Biswas:2011ar}, and show, although nonlocal, its quantum nature can still be evidenced. This means that the type of locality assumption we require in our experiment is not prohibitively restrictive and depends on the scale of non-locality of the theory.

\item{\bf Linearized gravity:} Note that we are always working in a regime of weak field gravity, linearized around the Minkowski background. In this way we avoid highly nontrivial space-times as the background (the experiment is to be carried out on Earth or a sattelite in space). This also means that the gravitational potential is always bounded below unity. In fact, below the millimetre scale we have no direct constraint on Newtonian $1/r$ potential~\cite{Kapner:2006si}~\footnote{Recently, the bound on short distance gravitational potential has been improved but the constraints are for the Yukawa type gravitational potential, which depends also on the strength of the Yukawa interaction~\cite{Kapner:2006si,Tan:2016vwu,Haddock:2017wav}. }. We are working in a regime of roughly $>100$ microns, and for the masses under consideration, the gravitational interaction is indeed weak and justifies the treatment of linearized gravity. At distances $>100$ microns the Casimir interaction is weaker than that of the gravitational interaction, see~\cite{Bose:2017nin}. We have also outlined in Ref.\cite{Bose:2017nin} how to get rid of all the competing electromagnetic forces, so that the only force is gravitational, we have to also ensure that no as yet unknown  ``fifth force'' acts here which essentially can also entangle the masses as a Newtonian force would do. This again, is easy to ensure for separations $>$ 100 micron for which Newtonian gravity has been very well tested. Similarly, the velocities of the masses are firmly in the non-relativistic regime so that the physics is well described by the Newtonian regime. 

\item{\bf Definition of a classical field: \label{item:classical definition}} Note that here we are {\em not} defining what makes something quantum, but rather  we clarify at the outset what we mean by a ``classical'' field. We simply take a classical field to be an entity which with general probabilities $P_j$ has fixed (unique) values  $h_{\mu \nu}^{j}$ at each point of space-time (here we have used a tensor field in the definition, but it could be scalar, spinor etc.). Of course a special case of that is when there are no probabilities at all -- the field just has a value $h_{\mu \nu}$. There is a reason that we are using a much broader definition than simply a unique value -- namely we are allowing also the probability of the field statistically having different values with different probabilities. This is just to carefully emphasize that the statistical nature of something does not make it quantum (think of a classical dice) -- quantum comes with the possibility to go beyond statistical mixtures of field configurations to coherent superpositions of field configurations. Additionally, we demand that a classical field means that we are not even allowed to think of a Hilbert space for the field, i.e., even joint quantum states of fields with other (say, matter) systems is disallowed, i.e., states of the form $\sum_j \sqrt{P_j} |j\rangle |h^{j}_{\mu \nu}\rangle$ are {\em not} allowed. Only allowed joint states of quantized matter and classical field are the Probability distributions $P_j$ of configurations $\{ |j\rangle\langle j|, h_{\mu \nu}^{j}\}$, where $h_{\mu \nu}^{j}$ is a tensor for each point in space time, but {\em not} an operator valued quantity. Our definition of classical field, and the consequences which follow from it (cf. Section \ref{sec:LOCC}) are very standard (if one expands significantly the definition of what is meant by classical, one will, of course, get other consequences \cite{hall2018two}). Thus we do not have a Here we are defining a classical field as, for example, used by Feynman during his 1957 debate with other researchers on the quantum nature of gravity \cite{feynman1957talk}  ``... if I have an amplitude for a field, that's what I would define as a quantized field.'' So a classical field is one which has probabilities for various field configurations rather than amplitudes for various field configurations.

\end{itemize}

\section[Quantum Gravity]{A Quantum Origin of the Newtonian Potential in Linearized Quantum Gravity}

\begin{figure}[h]
	\includegraphics[width=\columnwidth]{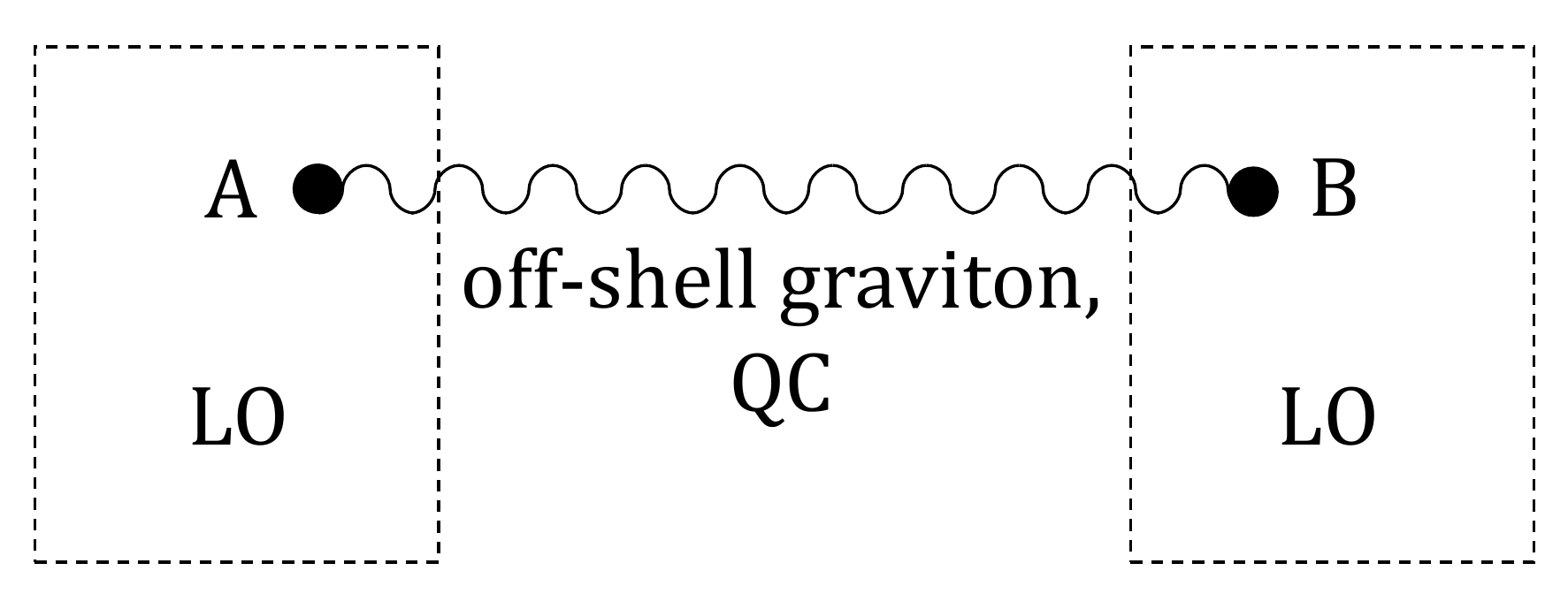}
	\caption{T-channel scattering Feynman diagram which is real with time in the vertical direction and space in the horizontal direction with zero momentum transfer. The dashed rectangle shows the Local Operation (LO) region during the interaction, where the particles A and B interact locally with the off-shell graviton. The curved line represents the exchange of off-shell gravitons which acts as a mediating Quantum Channel (QC). \label{fig:Feynman diagram}}
\end{figure}


The Einstein-Hilbert equation around the Minkowski background is given by:
\begin{equation}
S_{EH}=\frac{1}{4}\int d^4x\, h_{\mu\nu}\,\mathcal{O}^{\mu\nu\rho\sigma}\,h_{\rho\sigma}+\mathcal{O}(\kappa h^3),\label{eq:lin-action-newtonian}
\end{equation}
where $\mathcal{O}(\kappa h^3)$ takes into account of higher order terms in the perturbation, while the four-rank operator $\mathcal{O}^{\mu\nu\rho\sigma}$ is totally symmetric in all its indices and defined as
\begin{equation}
\begin{array}{ll}
\displaystyle \mathcal{O}^{\mu\nu\rho\sigma}:= \displaystyle \frac{1}{4}\left(\eta^{\mu\rho}\eta^{\nu\sigma}+\eta^{\mu\sigma}\eta^{\nu\rho}\right)\Box-\frac{1}{2}\eta^{\mu\nu}\eta^{\rho\sigma}\Box&\\
\,\,\,\,\,\,\,\,\,\,\,\,\,\,\,\,\,\displaystyle +\frac{1}{2}\left(\eta^{\mu\nu}\partial^{\rho}\partial^{\sigma}+\eta^{\rho\sigma}\partial^{\mu}\partial^{\nu}-\eta^{\mu\rho}\partial^{\nu}\partial^{\sigma}-\eta^{\mu\sigma}\partial^{\nu}\partial^{\rho}\right),&
\end{array}\label{4-rank-oper}
\end{equation}
for $\Box=\eta_{\mu\nu}\nabla^{\mu}\nabla^{\nu}$. By inverting the kinetic operator we obtain the graviton propagator around the Minkowski background, and its saturated between two conserved currents (in our case these are energy momentum tensors, see below Eq.(\ref{eq: exch})) and gauge independent part 
is given by \cite{VanNieuwenhuizen:1973fi,Biswas:2013kla}~\footnote{ By definition, off-shell graviton does not obey the classical equations of motion, see this discussion below. The graviton propagator in general relativity can be recast as:
$ \Pi_{\mu\nu\rho\sigma}(\vec k) =\frac{1}{2\vec k^2}\left(\eta_{\mu\rho}\eta_{\nu\sigma} +\eta_{\nu\rho}\eta_{\mu\sigma}-\eta_{\mu\nu}\eta_{\rho\sigma}\right)$.  This propagator can be obtained  either in a particular gauge known as harmonic gauge, or it can be obtained by using the projection operator technique defined in the Appendix-A. For the details of the projection operator, see Refs.~\cite{Rivers:1964,VanNieuwenhuizen:1973fi,Biswas:2013kla}.}

\begin{equation}
\Pi_{\mu\nu\rho\sigma}(k)=\left(\frac{\mathcal{P}_{\mu\nu\rho\sigma}^2}{k^2}-\frac{\mathcal{P}^0_{s,\,\mu\nu\rho\sigma}}{2k^2}\right),\label{propag}
\end{equation}
where $\mathcal{P}^2$ and $\mathcal{P}^0_s$ are two spin projection operators projecting along the spin-$2$ and spin-$0$ components, respectively; see Refs.\cite{Rivers:1964,VanNieuwenhuizen:1973fi,Biswas:2013kla} for further details. 
The Newtonian potential between the two masses, $T^{\mu\nu}_1\sim m\delta^{\mu}_{0}\delta^{\nu}_{0}\delta^{(3)}(\vec r)$ and the unit mass $T^{\mu\nu}_2\sim \delta^{\mu}_{0}\delta^{\nu}_{0}\delta^{(3)}(0)$ can be computed via a scattering diagram in quantum field theory. This can be envisaged by a transition amplitude in quantum mechanics, which we will discuss briefly. In fact, this part of the discussion is common to any quantum field theory, which has well defined initial and final states. Various examples will be Coulomb interaction via an exchange of a photon, or Yukawa potential via an exchange of a meson field, see \cite{thomson_2013}.
 
Note that in quantum mechanics the transition matrix element is given by the perturbation expansion~\footnote{There are many textbooks on quantum field theory which the readers can refer. Here we have provided a lucid discussion by Ref.\cite{thomson_2013}, see section 5.1 and 5.2.}: 
\begin{equation}\label{fn: scatting}
T_{fi}=\langle f|V| i \rangle+\sum _{j\neq i }\frac{\langle f |V|j \rangle \langle j|V| i \rangle}{E_{i}-E_{j}}+\cdots
\end{equation}
The transition matrix element determines the transition rate of any process going from initial state $i$ to final $f$.
 The first term in the perturbation series, $\langle f|V| i \rangle$, can be imagined as {\it scattering in a fixed potential}. Such a scattering is considered unsatisfactory because the transfer of momenta happen without any mediating field. Also, the force obtained from such a potential will lead to violation of special theory of relativity, immediate action-at-a distance. Nevertheless, this is purely a classical scattering in a fixed potential. The potential here could be Coulomb or gravitational or Yukawa potential. In this sense, the potential here is purely a classical concept. The transition matrix is $T_{fi}= \langle \psi_f |V(r)|\psi_i\rangle$, where $V(r)$ is the static potential for Coulomb, Yukawa  or gravitational.
  
 The second term in the series can be viewed as {\it scattering via an intermediate state} $j$. In quantum field theory interactions between particles always happen via an exchange of a mediator, which can be understood in time ordered perturbation theory. For a process $a+b\rightarrow c+d$ via an exchange of quanta $X$ will have two time ordered diagrams. Summing the matrix element for both the time ordered diagrams yield a Feynman propagator, see~\cite{thomson_2013}. An off-shell/virtual exchange of the mediator $X$ satisfies the conserved energy momentum tensor at the two vertices, but does not satisfy the classical on-shell equations of motion. By this we mean that the propagator does not satisfy the Einstein energy-momentum relationship and it is termed as off mass-shell, or off-shell/virtual. By definition a graviton propagator, see Eq.(\ref{propag}) is a non-classical entity, precisely because $k^2 \neq0$, 
 and in order to find the potential, we are integrating over all possible values of $k$, see the derivation below, 
 Eq.~(\ref{eq: exch}).
 The forces between particles now result from the transfer of the momentum carried by the exchanged spin-2 graviton, which has two off-shell propagating degrees of freedom as shown in Eq.(\ref{propag}).

 For a non-relativistic setup, we are only interested in $00$ components. The two conserved vertices will be the two masses, $T^{\mu\nu}_1\sim m\delta^{\mu}_{0}\delta^{\nu}_{0}\delta^{(3)}(\vec r)$, and the unit mass $T^{\mu\nu}_2\sim \delta^{\mu}_{0}\delta^{\nu}_{0}\delta^{(3)}(0)$. The non-relativistic potential will be given by integrating all the momenta of the off-shell graviton propagator $\Pi_{0000}$:
\begin{align}
\Phi(r)= &\displaystyle -\kappa^2  \int\frac{d^3|\vec{k}|}{(2\pi)^3}T_1^{00}(k)\Pi_{0000}(k)T_2^{00}(-k)e^{i\vec{k}\cdot\vec{r}} \nonumber\\
= & \displaystyle -\frac{\kappa^2 m}{2} \int\frac{d^3|\vec{k}|}{(2\pi)^3}\frac{1}{\vec{k}^2}e^{i\vec{k}\cdot(\vec{r})}=-\frac{G m}{r}\,,\label{eq: exch}
\end{align}
which recovers the Newtonian potential.
The above potential has been obtained in a scattering theory. 
 \footnote{The above potential result could have been obtained following real time formalism, or Schwinger-keldysh formalism~\cite{Schwinger:1960qe,Keldysh:1964ud}. This method is more powerful for doing out-of-equilibrium, or one(higher)-loop computations as well, but here we are interested in the tree level, non-relativistic, scattering diagram, for which the answer would be exactly the same as that of Schwinger-Keldysh formalism, see~\cite{park2010solving}. }. Note that we are integrating over the off-shell/virtual massless graviton. A virtual or off-shell particle is a mathematical construction, which represents the effect of summing over all possible time-ordered diagrams. By definition, such a process (summing over intermediate states $|j\rangle$ as in Eq.(\ref{fn: scatting})) involves quantum superposition of different off-shell graviton states -- making it a entirely quantum process. An off-shell particle does not satisfy the classical equations of motion, and therefore it is considered to be non-classical.

For different modifications of the graviton propagator the potential will be different,  for instance if there is an extra scalar degree of freedom propagating, then potential will be Yukawa type~\footnote{Original Yukawa potential was also obtained by the scattering amplitude of an exchange of a meson field between the two fermions. }. We will consider one such modification in the non-local setup. The analysis of this section, however, provides the quantum mechanism necessary to make sense of the result that the observation of entanglement generation mediated by gravity implies the quantum nature of linearized gravity. In other words:\\

{\it  If and only if gravity is quantum, then this would inevitably lead to entanglement between two or more, generic, matter states.}

\section[LOCC]{Impossibility of Entanglement Through a Classical Field\label{sec:LOCC}}

The aim here is to test whether $1/r$ potential is being mediated by a classical or a quantum channel. One can also obtain $1/r$ potential without a mediator, where there is no need to invoke a graviton as a propagator or a mediator. In this case, the particles act as sources for field which gives rise a potential in which other particles scatter.  This is precisely the classical scenario.
This classical mediator could be a potential $V$ as presented by the first term in Footnote~\ref{fn: scatting}. We will here show that with such a set-up it is \emph{impossible} to develop entanglement as this may not be immediately clear to all readers. Those familiar with the topic will recognise this as the well established entanglement non-increasing property of Local Operations and Classical Communication (LOCC)~\cite{bennett1999quantum}. Consider the two quantum bodies $A$ and $B$ to be initially in the separable state $|\psi\rangle_{A}\otimes|\phi\rangle_{B}$. These are acted on by the set of local operators $\left\{\hat{A}_{i,j}\right\}$ and $\left\{\hat{B}_{j,k}\right\}$ respectively. These could be enacted by experimentalists Alice and Bob, can occur due to natural evolution of the systems in isolation, or as a direct result of interacting with the shared classical channel (field). Here the labels $i$ and $k$ allow for muliple operators acting on each body, while we allow for a classical channel to transmit arbitrary classical information, here encoded in the parameter $j$ {\footnote{The parameter $j$ can encode any classical information, for example a classical metric perturbation $h_{\mu\nu}$ as created by the mass distributions of $A$ and $B$ acting as sources.  In the case of semi-classical gravity, this $h_{\mu\nu}$ could be a function of the expectation value of the source stress energy tensor, i.e. $\left\langle T_{\mu\nu}\right\rangle$, such as in the Schroedinger-Newton equation, or due to the source mass after its been localised by stochastic, spontaneous collapse as is predicted by collapse models ~\cite{tilloy2016sourcing} or something else entirely.}. In the case of a classical gravitational field this could be some function of the average over the position distributions of $|\psi\rangle_{A}$ and $|\phi\rangle_{B}$ as used in the Schroedinger-Newton equation, or it could encode the result of the stochastic collapse of the wavefunctions $|\psi\rangle_{A}$ and $|\phi\rangle_{B}$ or something else entirely. To account for this generality, it is necessary to use the density matrix formalism to describe the evolution of the masses. Furthermore we can write the evolution of each quantum state by some total evolution operator $\hat{A}_{i,j}(t)\otimes\hat{B}_{j,k}(t)$. In this way we can write the arbitrary evolution of the local operations acting on the two quantum masses with arbitrary classical information shared between them as

\begin{align}
\rho\left(t\right)=& \sum_{i}\sum_{j}\sum_{k} p\left(i\right)p\left(j\right)p\left(k\right) \nonumber\\
&\times\hat{A}_{i,j}(t)|\psi\rangle_{A}\langle\psi|_{A}\hat{A}^{\dagger}_{i,j}(t)\otimes\hat{B}_{j,k}(t)|\phi\rangle_{B}\langle\phi|_{B}\hat{B}^{\dagger}_{j,k}(t) \label{eq:LOCC}
\end{align}

where $p\left(i\right)$ and $p\left(k\right)$ encode probabilities for various operators acting on the matter states and $p\left(j\right)$ can encode the classical probabilities corresponding to different classical field configurations $h_{\mu \nu}^{j}$ present (multiple field configurations leading to different values of the parameter $j$ can occur if there are any stochastic collapses of the gravitational field due to any stochastic collapses of the matter states). Eq.~\ref{eq:LOCC} can reproduce arbitrary local evolution of both quantum masses, and arbitrary \emph{classical} correlations between the two masses. It is however trivial to see that regardless of this the total state remains separable, that is, unentangled. The use of the total evolution operators hides much of the details of how the quantum states evolve, for example, for physically realistic evolutions one should expect them to include the necessary provisions to maintain causality. This hidden detail however \emph{cannot} change the final conclusion, that the exchange of classical information, be it via a gravitational field or telephone wire, coupled with arbitrary local operations will not entangle two quantum systems.

Because of the generality of the above treatment, it follows that starting from a separable state of two systems $A$ and $B$, quantum entanglement cannot be generated by any model in which gravity is a classical field (classical according to the standard definition given in section~\ref{item:classical definition}). This automatically encompasses all specific models such as the Moller-Rosenfeld semiclassical gravity model or models where the matter field undergoes collapses and sources a stochastic classical gravitational field~\cite{tilloy2019does,derakhshani2016probing}.

\section{Entanglement in Gravitationally Interacting Interferometer}
\label{sec:Interferometer}

For completeness, in this section will provide an overview of the experiment being discussed throughout this paper. The set-up, shown in Fig.~\ref{fig:Experiment diagram}, consists of two mesoscopic mass ($\sim10^{-14}$kg) microspheres with embedded spins traversing two Stern-Gerlach interferometers in close proximity to one another. The two masses become entangled due to the varying gravitational interaction between them due to the differing separations of the interferometer arms. The interferometric process is completed by bringing together the two spatial wavepackets which leads to the path phase differences being imprinted into the particles spin state, with any entanglement measured by their spin correlations (cf Sec.~\ref{sec:Entanglement Witness}).

\begin{figure}
	\includegraphics[width=\columnwidth]{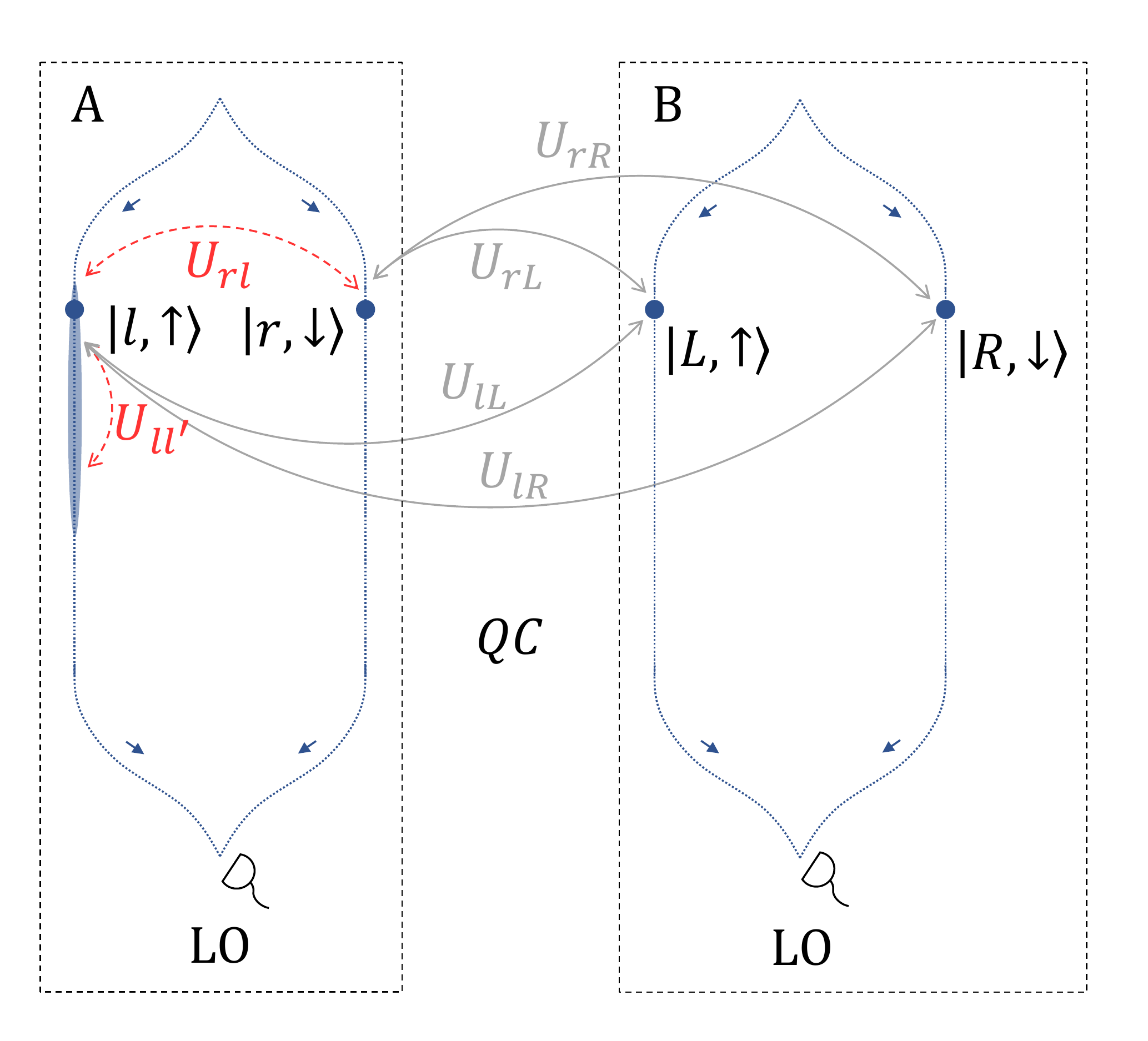}
	\caption{Experiment set-up showing the two interferometers, the two particles ($A$ and $B$), their trajectories (dotted blue path) and their corresponding position and equivalently spin state, and the quantum channel (QC) mediating the interactions between the four position states. The dashed rectangle encompasses the local operations (LO) regions for particles $A$ and $B$. The solid grey lines show the gravitational interactions which lead to entanglement, while the dashed red lines are example of some of the unwanted interactions which could occur for non-Fock mass states.  \label{fig:Experiment diagram}}
\end{figure}

Each mass will initially be in a spatial superposition of being both `left' and `right' with the two particle joint state as a function of the form $\left|ab\right\rangle$ where $a\in\left\{l,r\right\}$, $b\in\left\{L,R\right\}$ as shown in Fig.\ref{fig:Experiment diagram}.  The two masses are treated as non-relativistic (stationary) point particles, both with mass $m$ such that the only non-zero component of the stress energy tensor will be
\begin{equation}
	T^{00}=m\delta^3\left(\vec{x}-\vec{x}_a\right) +m\delta^3\left(\vec{x}-\vec{x}_b\right) \label{eq: mass-energy tensor rep}
\end{equation}
\footnote{It is perhaps worth clarifying that when the mass exists in a spatial superposition of being in two locations  (i.e. superposition of $\vec{x}_a=\vec{x}_l$ and $\vec{x}_a=\vec{x}_r$ and similar for $\vec{x}_b$), we \emph{do not} have $T^{00}=\frac{1}{2}\left(m\delta^3\left(\vec{x}-\vec{x}_l\right)+m\delta^3\left(\vec{x}-\vec{x}_r\right)\right)+\frac{1}{2}\left(m\delta^3\left(\vec{x}-\vec{x}_L\right)+m\delta^3\left(\vec{x}-\vec{x}_R\right)\right)$.} To model the results of interactions between two particles, both in superposition states, we employ the Feynman style logic treating the resulting total state as the sum of four individual amplitudes, each belonging to the separate field configurations created by each possible joint state for the matter, with each component evolving as
\begin{equation}
	\left|ab\right\rangle \rightarrow e^{-i\frac{Gm^2\tau}{\hbar r_{ab}}}\left|ab\right\rangle \label{eq:newtonain interaction}
\end{equation}
where this evolution is derived from Eq.~\ref{eq: exch} and $\tau$ is the interaction time, implicitly assuming each mass is within the light cone of the other situated with its origin at the point in which the superposition is created ($t=0$). Using Eq.~\ref{eq:newtonain interaction}, and considering the four interactions shown as solid grey lines in Fig~\ref{fig:Experiment diagram} gives

\begin{eqnarray}
\left|\psi\left(0\le t<\delta t\right)\right\rangle =&\frac{1}{\sqrt{2}}\left(\left|l\right\rangle + \left|r\right\rangle \right)\otimes\frac{1}{\sqrt{2}}\left(\left|L\right\rangle + \left|R\right\rangle \right),~ \\
\left|\psi\left(t=\tau+\delta t\right)\right\rangle =&\frac{1}{2}\bigg(e^{-i\frac{Gm^2\tau}{\hbar r_{lL}}}\left|lL\right\rangle + e^{-i\frac{Gm^2\tau}{\hbar r_{lR}}}\left|lR\right\rangle \nonumber\\
& + e^{-i\frac{Gm^2\tau}{\hbar r_{rL}}}\left|rL\right\rangle + e^{-i\frac{Gm^2\tau}{\hbar r_{rR}}}\left|rR\right\rangle\bigg)~~~ \label{eq:general form of final state}
\end{eqnarray}
giving the entanglement between the two masses as found as in~\cite{Bose:2017nin}, where $r_{ab}=\left|\vec{x}_a-\vec{x}_b\right|$ is the distance between the two masses. As such we have the standard Newtonian potential appearing to mediate the interaction between the two masses.

\textit{NOON States:} In quantum field theory the masses should be considered as excitations of a quantum field~\cite{blencowe2013effective}, such as a fock state of fields $\hat{\phi}_1$ and $\hat{\phi}_2$ where
\begin{equation}
\left|ab\right\rangle=\left(\hat{\phi}_1^{\dagger}\left(a\right)\hat{\phi}^{\dagger}_2\left(b\right)\right)\left|0\right\rangle. \label{eq:explaining state notation}
\end{equation}
Here $\hat{\phi}_{i}^{\dagger}\left(x\right)$ is the creation operator which creates a mass centred at $x$ and where each object is in a spatial superposition
\begin{equation}
\frac{1}{\sqrt{2}}\left(\hat{\phi}^{\dagger}_1\left(l\right)+\hat{\phi}^{\dagger}_1\left(r\right)\right)\left|0\right\rangle \label{eq:fock state superposition}
\end{equation}
such that there cannot be any interaction between the two arms within an interferometer of the form shown in Fig~\ref{fig:Experiment diagram} by $U_{lr}$. For each mass, the mass field must be in a state qualitatively similar to Eq.~\ref{eq:fock state superposition}.
In the proposals the mass states are taken to be in this exact state. One can also identify each mass as fundamental or composite. Rather than a single object, a collection of fundamental particles in a NOON state ($\left|n,0\right\rangle+\left|0,n\right\rangle$ in the Fock basis), which corresponds to a superposition of $n$ fundamental particles (nucleons, electrons etc.) in the first arm of the interferometer and $0$ in the second and vice versa.  Furthermore, it is sufficient to consider a single excitation of a large ($10^{-14}$ kg) mass field as: (a) there is not enough energy to create a second $10^{-14}$ kg excitation of the mass, and (b) not enough energy to disassociate the mass into its individual components. As such the internal dynamics of the mass is unimportant here. Also, if a coherent state in each arm of the interferometer of the form $e^{\frac{\left|\alpha_l\right|^2}{2}}e^{\frac{\left|\alpha_r\right|^2}{2}}e^{\alpha_l\hat{\phi}^{\dagger}_1\left(l\right)}e^{\alpha_r\hat{\phi}^{\dagger}_1\left(r\right)}\left|0\right\rangle$ was used, as one would expect in a non/weakly-interacting Bose-Einstein condensate (BEC), this would create interactions of the form $U_{lr}$ which will not result in entanglement of the form necessary to demonstrate the quantum nature of gravity. 
  
 If a continuous stream of particles were used, such interactions ($U_{aa'/bb'}$ and $U_{lr/LR}$) could dominate any signal from the inter-arm interactions ($U_{aB}$), effectively overwhelming the entangling signal in the noise of these other interactions. For this reason NOON states of BECs would have to be used~\cite{howl2019exploring}. For example, consider if the mass state employed was $\left|\psi\left(0\le t<\delta t\right)\right\rangle =\hat{\phi}^{\dagger}_1\left(l\right)\hat{\phi}^{\dagger}_1\left(r\right)\hat{\phi}^{\dagger}_2\left(L\right)\hat{\phi}^{\dagger}_2\left(R\right)\left|0\right\rangle$, then interactions of the form $U_{lr}$ would be allowed, and Eq.~\ref{eq:general form of final state} would become
\begin{eqnarray}
\left|\psi\left(t=\tau+\delta t\right)\right\rangle =&e^{-i\frac{Gm^2\tau}{\hbar}\left(\frac{1}{r_{lL}}+\frac{1}{r_{lR}}+\frac{1}{r_{rL}}+\frac{1}{r_{rR}}+\frac{1}{r_{lr}}+\frac{1}{r_{LR}}\right)} \nonumber\\
&\times\hat{\phi}^{\dagger}_1\left(l\right)\hat{\phi}^{\dagger}_1\left(r\right)\hat{\phi}^{\dagger}_2\left(L\right)\hat{\phi}^{\dagger}_2\left(R\right)\left|0\right\rangle\label{eq:incorrect form of final state}
\end{eqnarray}
which is not an entangled state. Thus is it necessary to prepare the matter states in NOON states of a quantum field during the initialisation of the experiment.

\section{Witnessing Entanglement through Measurement Statistics \label{sec:Entanglement Witness}}

The experimental proposal~\cite{Bose:2017nin} will results in an output state consisting of two entangled spin qubits (that is of course assuming gravity is quantum). To understand how such entanglement is verified, it is worth discussing what quantum entanglement is. For a bipartite state to be entangled means the state cannot be written as the tensor product of the states of each particles, that is, a state is not entangled (it is separable) if it can be written
\begin{equation}
\rho=\sum_j p\left(j\right)\rho_{A,j}\otimes\rho_{B,j}
\end{equation}
where $\left|\phi\right\rangle_A$ and $\left|\chi\right\rangle_B$ are arbitrary states belonging to the Hilbert space of particles A and B respectively and $\sum_j p\left(j\right)=1$. If we restrict ourselves to bipartite, pure qubit states, then we can understand and quantify entanglement by the Von Neumann entropy of the reduced density matrix, defined as

\begin{equation}
	\mathcal{S}\left(\hat{\rho}_A\right)=-Tr\left[\hat{\rho}_A\log\left(\hat{\rho}_A\right)\right]. \label{eq:Entanglement entropy}
\end{equation}

Take for example the maximally entangled, product state
\begin{equation}
	\left|\psi\right\rangle=\frac{1}{\sqrt{2}}\left(\left|00\right\rangle+\left|11\right\rangle\right) \label{eq:eg maximally entangled state}
\end{equation}

then we have a corresponding density matrix $\hat{\rho}=\left|\psi\right\rangle\left\langle\psi\right|$. Tracing out one of the particles leaves a reduced density matrix
\begin{equation}
	\hat{\rho}_A=Tr_B\left(\hat{\rho}\right)\propto\mathbb{I}_A
\end{equation}
which corresponds to a maximal entropy state, where $\mathcal{S}\left(\hat{\rho}_A\right)=1$. This can be understood as fully entangled particles will contain information about the other particle too, by throwing away the information held by only one of the particles (tracing it out), the result contains no useful information. If the initial state was instead separable, then the reduced density matrix would correspond to that for a completely ordered state. In view of the above, one might expect witnessing the masses initially in pure states (low entropy) evolving into mixed states (high entropy) would prove entanglement, however this is not the case. In a realistic experiment, decoherence (such as entanglement with the environment) which create mixed state, and so also maximise entropy, cannot be ruled out. As such no conclusion could be drawn from actual measurements of the entropy.

Alternatively entanglement measures which are compatible with mixed states can be used, for example concurrence or an entanglement witness can be used. The concurrence can be calculated for a general (pure or mixed) two qubit state, which the two spin states can be thought of as, using

\begin{equation}
	C\left(\rho\right)=\textrm{max}\left\{0,\lambda_1-\lambda_2-\lambda_3-\lambda_4\right\}
\end{equation}
where $\lambda_i$ is the square root of the eigenvalues of the matrix $\rho\tilde{\rho}$ arranged in decreasing order, for $\tilde{\rho}=\left(\sigma_y\otimes\sigma_y\right)\rho^*\left(\sigma_y\otimes\sigma_y\right)$. Again, this is maximised by maximally entangled states such as Eq.~\ref{eq:eg maximally entangled state}, which gives $C\left(\rho\right)=1$. However, to calculate the concurrence the entire states density operator is needed which requires full state tomography, a measurement intensive process (requires 6 expectation and 9 correlation measurements). To avoid this entanglement witnesses can be used, which looks at correlations between the two particles, in this way any measured entanglement is confirmed to be between the two particles and not one particle and its environment. Such an entanglement witness $\mathcal{W}\left(\hat{\rho}\right)$ is defined such that it has the property that it evaluates to greater than $1$ only if $\hat{\rho}$ is entangled. It is important to note that the converse is not true, that is,  if it is not greater than $1$, it does not imply anything about $\hat{\rho}$. Furthermore such witnesses need to be created to detect the specific entangled state which can be difficult in general, and will necessarily detect different a state as entangled, even if it is maximally entangled. However, due to the simple nature of the final state, a suitable witness was found to be
\begin{equation}
\mathcal{W}=\left|\left\langle\sigma_x^{\left(1\right)}\otimes\sigma_z^{\left(2\right)}\right\rangle-\left\langle\sigma_y^{\left(1\right)}\otimes\sigma_y^{\left(2\right)}\right\rangle\right|\,,
\end{equation}
which is sufficient for discriminating the entanglement as it is expected to develop in the tabletop experiment. It also only requires two sets of measurements for each particle

\section{non-local Gravity}

The entanglement experiment protocol is also not limited to probing the quantum nature of local gravitational models, it could also be used to probe the quantum nature of gravity which is non-local over a microscopic scale as well as modifications to the gravitational potential at short distances. For instance, modifications of gravity in the ultraviolet. The most general quadratic action in $4$ dimensions, which is invariant under parity and also torsion-free is given by \cite{Biswas:2011ar}

\begin{equation}
\begin{array}{rl}
S=& \displaystyle \frac{1}{16\pi G}\int d^4x\sqrt{-g}\left\lbrace \mathcal{R}+\beta\left(\mathcal{R}\mathcal{F}_1(\Box_s)\mathcal{R}\right.\right.\\[3mm]
& \displaystyle  \,\,\,\,\,\,\,\,\,\,\,\,\,\,\,\,\left.\left.+\mathcal{R}_{\mu\nu}\mathcal{F}_2(\Box_s)\mathcal{R}^{\mu\nu}+\mathcal{R}_{\mu\nu\rho\sigma}\mathcal{F}_3(\Box_s)\mathcal{R}^{\mu\nu\rho\sigma}\right)\right\rbrace,
\end{array}
\label{quad-action}
\end{equation}
where $\Box_s=\Box/M_s^2$ and $M_s$ is considered as the fundamental scale of non-locality, which in the context of string theory corresponds to the string scale. Within $M_s^{-1}$ the micro-causality is violated~\cite{Efimov:1967pjn,Tomboulis:1997gg,Tomboulis:2015gfa,Buoninfante:2018mre}. For $\Box\ll M_s^2$, the theory becomes that of a local theory with a low energy limit given purely by the Einstein-Hilbert action~\cite{Biswas:2011ar}. Furthermore, by considering such a modified gravity we are also demonstrating that our local gravity assumption is not as strict as it might appear provided the locality is violated at a microscopic level and the time and length scale of our experimental set up is larger than $M_s^{-1}$. The three gravitational form-factors $\mathcal{F}_{i}(\Box_s)$ are covariant functions of the d'Alembertian and can be uniquely determined around the Minkowski background~\cite{Biswas:2011ar,Biswas:2013cha}. We can set $\mathcal{F}_3(\Box_s)=0,$  without loss of generality up to quadratic order in the metric perturbation around the flat background, we can keep the massless spin-$2$ graviton as the only dynamical degree of freedom by imposing the following condition~\footnote{In this paper we will only consider analytic form-factors. However, it is worth mentioning that non-local models with non-analytic differential operators have been investigated by many authors; see, for example, Refs. \cite{Barvinsky:1985an,Deser:2007jk,Conroy:2014eja}.}:
$2\mathcal{F}_1(\Box_s)=-\mathcal{F}_2(\Box_s)$ as shown in Ref.\cite{Biswas:2011ar} around the Minkowski background.
By expanding around Minkowski, $g_{\mu\nu}=\eta_{\mu\nu}+\kappa h_{\mu\nu},$ we obtain
\begin{equation}
S=\frac{1}{4}\int d^4x\, h_{\mu\nu}\,(1-\mathcal{F}_1(\Box_s)\Box_s)\mathcal{O}^{\mu\nu\rho\sigma}\,h_{\rho\sigma}+\mathcal{O}(\kappa h^3),\label{eq:lin-action-IDG}
\end{equation}
and the saturated and gauge independent part of the propagator is given by \cite{Biswas:2011ar,Biswas:2013kla}
\begin{equation}
\Pi_{\mu\nu\rho\sigma}(k)=\frac{1}{1+\mathcal{F}_1(k)k^2/M_s^2}\left(\frac{\mathcal{P}_{\mu\nu\rho\sigma}^2}{k^2}-\frac{\mathcal{P}^0_{s,\,\mu\nu\rho\sigma}}{2k^2}\right),\label{propag-1}
\end{equation}
where $\mathcal{P}^2/k^2-\mathcal{P}^0_{s}/2k^2$ is the graviton propagator of Einstein's GR, see Eq.(\ref{propag}).  Note that in order not to introduce any extra dynamical degrees of freedom other than the massless spin-$2$ graviton, we need to require that the function $1+\mathcal{F}_1(k)k^2/M_s^2$ does not have any zeros, i.e. that it is an {\it exponential of an entire function}\cite{Biswas:2011ar}:
\begin{equation}
1+\mathcal{F}_1(k)\frac{k^2}{M_s^2}=e^{\gamma(k^2/M_s^2)},\label{choice}
\end{equation}
where the $\gamma(k^2/M_s^2)$ is an entire function. We will mainly work with the simplest choice $\gamma(k^2)=k^2/M_s^2$,  see also Ref.~\cite{Frolov:2015bia,Frolov:2015bia,Edholm:2016hbt} for other examples of entire functions. In all these examples the short distance behaviour becomes soft and in the IR the gravitational potential matches that of Newtonian prediction. Now we can compute the scattering diagram. The key difference from a local gravitational theory is that now the existence of a new scale, $M_s$, which determines the interaction at short distances. For $k^2\ll M_s^2$, the non-local contribution becomes exponentially small, or in length scale $r> M_s^{-1}$, the theory predicts the results of local Einstein-Hilbert action.

We can now compute the gravitational potential by integrating all the momenta of the off-shell graviton, assuming the two vertices are non-relativistic. Essentially, taking the $T^{00}$ components only, and with modified graviton propagator, we obtain:
\begin{equation}
\begin{array}{rl}
\Phi_{IDG}(r)=&\displaystyle -\kappa^2  \int\frac{d^3|\vec{k}|}{(2\pi)^3}T_1^{00}(k)\Pi_{0000}(k)T_2^{00}(-k)e^{i\vec{k}\cdot(\vec{r})}\\
= & \displaystyle -\frac{\kappa^2 m}{2} \int\frac{d^3|\vec{k}|}{(2\pi)^3}\frac{e^{-\vec{k}^2/M_{s}^2}}{\vec{k}^2}e^{i\vec{k}\cdot(\vec{r})},\\
=& \displaystyle -\frac{G m}{r}{\rm Erf}\left(\frac{M_{s}r}{2}\right)\,.
\end{array}\label{exch}
\end{equation}
Note that the gravitational potential is now modified.

\begin{figure}
	\includegraphics[width=\columnwidth]{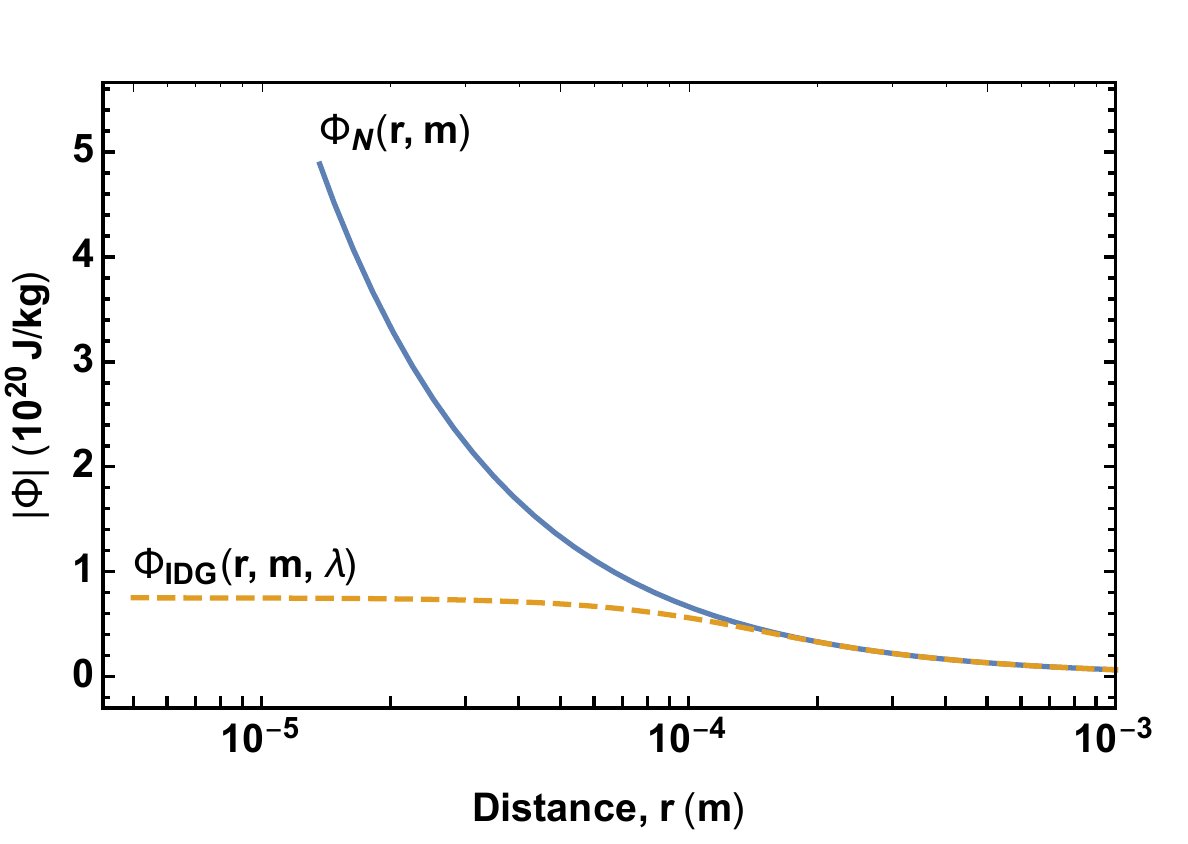}
	\caption{Potential energy per unit test mass as generated by a $m=10^{-14}$ kg source mass for both the standard newtonian potential ($\Phi_{N}$) and the modified infinite derivative gravity potential ($\Phi_{IDG}$). The non-local parameter for $\Phi_{IDG}$ was set to $M_s =0.004$ which corresponds to a non-local range $\lambda=5\times10^{-5}$ m, see~\cite{Edholm:2016hbt}.  \label{fig:Potential Energy}}
\end{figure}

In particular when $r<2/M_s$, the error function increases linearly with $r$, which cancels the denominator. Therefore at short distances, for $r<2/M_s$, the gravitational potential becomes constant and given by:
\begin{equation}
\Phi_{IDG}(r)\sim \frac{GmM_s}{\sqrt{\pi}}\,,
\end{equation}
while for $r> 2/M_s$, the error function approaches $\pm 1$, and therefore the potential recovers the standard newtonian potential, $-Gm/r$, as seen in Fig~\ref{fig:Potential Energy}.

One can compute various gravitational invariances, including the Kretschmann invariant, which remains constant as $r\rightarrow 0$, see~\cite{Buoninfante:2018xiw}. Indeed, note that this computation has been performed in the linear theory. To be consistent here, the gravitational singularity is ameliorated when the gravitational potential is still within the linear regime.

\begin{equation}
{2|\Phi_{IDG}(r)|}<1,~~mM_s < M_p^2\,,
\end{equation}

Since the entanglement phase depends on the potential, at short distances ($r<2/M_s$) the gravitational potential approaches constant as long as the inter separation distance is well within the non-local region.  It has also been shown that non-locality never exceeds beyond the non-local scale of $M_s$, see for instance~\cite{Pais:1950za,Efimov:1967pjn,Tomboulis:2015gfa,Buoninfante:2018mre}. Therefore, if all superposition components of the two masses are well inside the radius of $r= 2/M_s$, the entanglement phase, which is dependent on the potential varying for different spin components, will linearly go to zero. This has indeed very intriguing repercussions for the entanglement phase, despite the fact that the treatment of the linearized graviton remains quantum. The non-local interaction weakens the gravitational potential by smoothening out the spacetime. This serves as an interesting example how non-local interactions can alter the quantum behaviour of the many body system. However, for $r> 2/M_s$, the entanglement phase is the same as that of general relativity, which is similar to our previous local case.

The entanglement witness experiment results can be quantified by the two parameters $\Delta\phi_{LR}$ and $\Delta\phi_{RL}$ which we can compare for the two gravitational potentials considered here. For an experimental set-up involving $10^{-14}$ kg masses, $2.5\times10^{-4}$ m superpositions and a minimum separation of $2\times10^{-4}$ m, assuming standard Newtonian gravity, $\Delta\phi_{LR}=-0.125$ rad and $\Delta\phi_{RL}=0.439$ rad, whereas for IDG $\Delta\phi_{LR}=-0.125$ rad and $\Delta\phi_{RL}=0.435$ rad, for $M_{s}=0.004$ eV, which corresponds to $5\times 10^{-6}$m. This translates to an expected entanglement witness value $\mathcal{W}=1.223$ with IDG compared to $\mathcal{W}=1.224$ for standard Newtonian gravity. 

Given the power of entanglement entropy in fully quantifying the amount of entanglement in pure states, and its current importance in quantifying entanglement in quantum field theories~\cite{calabrese2004entanglement}, it can also be insightful to consider. Furthermore, although incredibly difficult, if we could ensure that the two-mass state, and eventually the two spin state to which the entanglement is mapped, remains pure, we can measure the full density matrix for one of the qubits with only 3 spin measurement settings and from that calculate the entanglement entropy given by Eq.~\ref{eq:Entanglement entropy}. The entanglement entropy for the experiment, given by
\begin{equation}
	\mathcal{S}\left(\hat{\rho}_A\right)=-\left(\lambda_{-}\log_{2}\left(\lambda_{-}\right)+\lambda_{+}\log_{2}\left(\lambda_{+}\right)\right)
\end{equation}
where 
\begin{align}
\lambda_{\pm}=&\frac{1}{2}\pm\frac{1}{2}\bigg[\frac{1}{2}\bigg(1+\cos\bigg(\frac{m\tau}{\hbar}(\Phi\left(r_0-\Delta x\right)\nonumber\\
&+\Phi\left(r_0+\Delta x\right)-2\Phi\left(r_0\right))\bigg)\bigg)\bigg]^{1/2},
\end{align}
$r_0$ and $\Delta x$ are the distance between the centre of the interferometers and superposition size respectively, is shown in Fig.~\ref{fig:Entropy} for both gravitational potentials. See Appendix~\ref{app:entropy} for more detail. In the experimental proposal, using time $\tau\approx2.5$ s we can see, although it is small, there is a quantitative difference between the two gravitational potentials with $\mathcal{S}\left(\hat{\rho}_A\right)=0.054$ for a Newtonian potential and $\mathcal{S}\left(\hat{\rho}_A\right)=0.053$ for IDG. The figure also shows that there is very little entanglement in the output state, which tends to zero for increasing separations, which is a result of the weakness and spatial dependence of gravity.
\begin{figure}
	\includegraphics[width=\columnwidth]{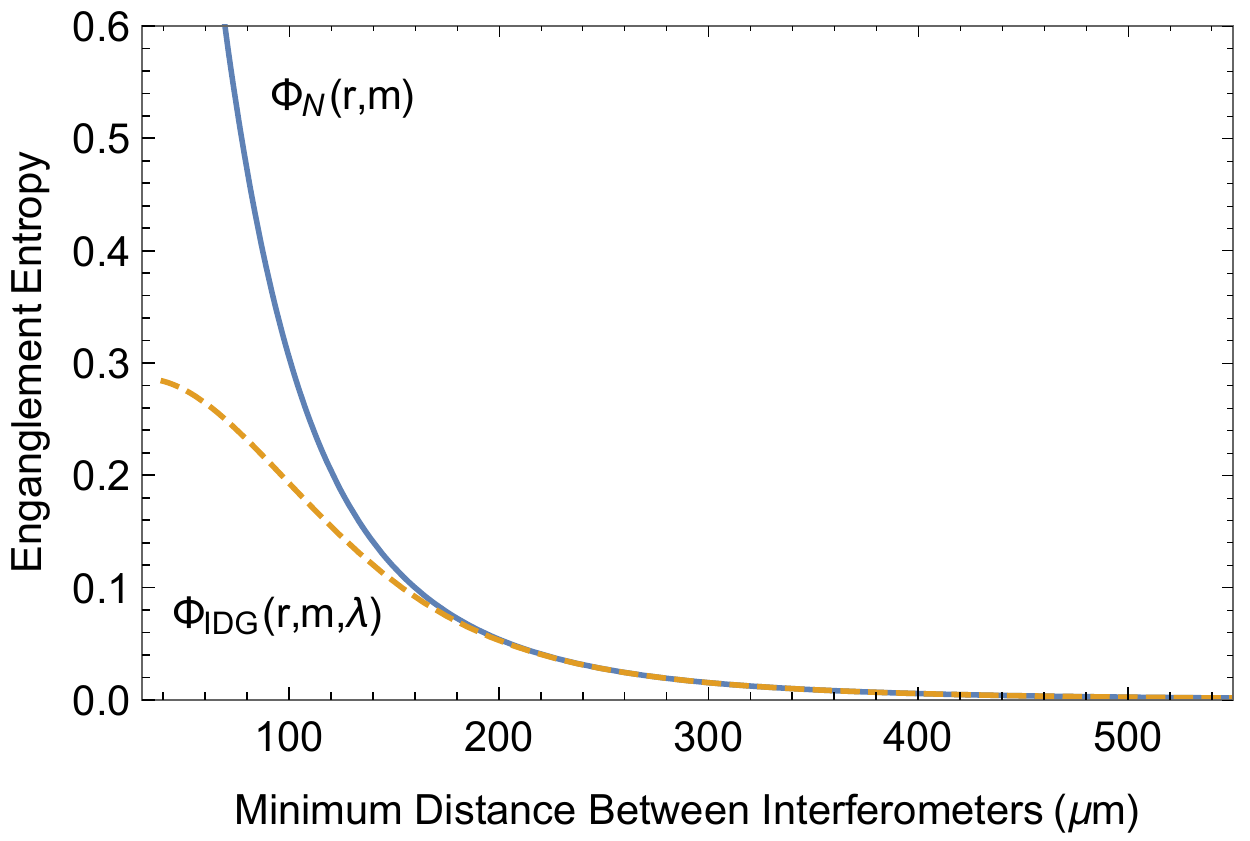}
	\caption{Entropy growth with minimum interferometer separation $\left(r_0-\Delta x\right)$, for both the standard newtonian potential ($\Phi_{N}$) and the modified infinite derivative gravity potential ($\Phi_{IDG}$). The non-local parameter for $\Phi_{IDG}$ was set to $M_s =0.004$ eV which corresponds to a non-local range $\lambda=5\times10^{-5}$ m, see~\cite{Edholm:2016hbt}. All other parameters match those provided in the original experimental proposal~\cite{Bose:2017nin}. \label{fig:Entropy}}
\end{figure}
As such there would be slight changes (revealing the scale $M_s$) in the result however, as the experiment is conducted outside the non-local region, all conclusions still hold, even in presence of non-local gravitational interaction. 

The above discussion provides also a way to probe short distance nature of gravity. Just from the current constraints on Newtonian $1/r$ potential, the direct experiments reveal that below millimetre distances $1/r$ potential is not constrained at all~\cite{Adelberger:2006dh,Tan:2016vwu}. Even the Yukawa-type potential between two neutral masses are constrained up to micro-meters. These experiments can directly place a constraint on the scale of non-locality to be of the order of $0.004$~eV~\cite{Edholm:2016hbt}. Our experimental protocol can in principle probe the nature of short distance gravity and the above example of non-local gravity illustrates an example of that. Besides the experimental query, non-local gravity also illustrates how entanglement entropy behaves in two distinct classes of theories, one were the $1/r$ singularity is present and the other where it is absent.

\section{Conclusion}
In this paper we have highlighted the {\it key} assumptions made in the paper \cite{Bose:2017nin}, in order to clarify what is meant by the statement that witnessing entanglement in the proposed experiment verifies the quantum nature of the gravitational field. First, we have presented the manner in which General Relativity lends itself to be quantized in a linearized limit. Doing so predicts the existence of gravitons and its minimal coupling to matter, including the off-shell gravitons, the exchange of which leads to the Newtonian gravitational force. Moreover, to generate entanglement through the quantum mediator one also requires the linearity of superpositions as highlighted in Eq.~\ref{eq:general form of final state}. It is only with the off-shell graviton (quantum) source for the Newtonian potential interaction \emph{and} the matter itself in a superposition that entanglement can be generated. Furthermore through the premise of LOCC (as clarified in Sec.~\ref{sec:LOCC}, for quantum masses sourcing a classical mediating field),  we know that the mediating channel, i.e. the gravitational field, \emph{cannot} be classical for the formation of entanglement. Further, the fact that the matter states can be described as a quantized field has been clarified, including that in this case these are in superpositions of Fock states and, more appropriately, when one considers the microscopic constituents, in NOON states.  As microcausality (cf. footnote~\ref{footnote:causality} in Sec.~\ref{sec:Assumptions}) is built into the standard relativistic quantisation, within which the virtual graviton exchange process acts, here the question of whether the Newtonian force is fundamentally action at a distance does not arise. As long as the masses are within each other's light cone, the potential is given by Eq.~\ref{eq: exch} and vanishes outside it.

We have also provided an example of non-local ghost free theory of gravity, where the gravitational potential is modified drastically to resolve the $1/r$ singularity. In this scenario, the gravitational interaction with matter becomes non-local, and provides a different prediction for the entanglement phase inside the non-local regime. Since, the experiment is always conducted outside the non-local region no significant change would be expected and this highlights that all our conclusions can still hold, even after breaking the local gravity assumption. Also of interest is the fact that the entanglement entropy, arising from local/non-local gravity, can be determined by the proposed experiment through measurements of the final spin states. A key observation of this paper is that if {\it gravity is quantized, the quantum matter degrees of freedom are generically in entangled states.} This fundamental and universal entanglement owes to the bare quantum nature of gravity, which remains finite  in spite of any other Standard Model like interactions we can imagine.

\begin{acknowledgments}
	AM's research is funded by the Netherlands Organisation for Scientific Research (NWO) grant number 680-91-119. SB would like to acknowledge EPSRC grants No. EP/N031105/1 and EP/S000267/1.
\end{acknowledgments}

\bibliography{bibliography}

\appendix
\section{Graviton propagator from spin projection operators \label{app:Propagator}}

We can expand Riemann tensor, Ricci tensor and Ricci scalar in up to order ${\cal O}(h)$
\begin{eqnarray}\label{ident-0}
R_{\mu\nu\lambda\sigma}=\frac{1}{2}(\partial_{[\lambda}\partial_{\nu}h_{\mu\sigma]}-\partial_{[\lambda}\partial_{\mu}h_{\nu\sigma]})\nonumber \\
R_{\mu\nu}=\frac{1}{2}(\partial_{\sigma}\partial_{(\nu}\partial^{\sigma}_{\mu)}-\partial_{\mu}\partial_{\nu}h-\Box
h_{\mu\nu})\nonumber \\
R=\partial_{\mu}\partial_{\nu}h^{\mu\nu}-\Box h .
\end{eqnarray}
Here we will study the full action Eq.~(\ref{quad-action}), which can be reduced to pure Einstein Hilbert action
as a low energy limit,  when we take $M_s \rightarrow \infty$. In this regard our treatment will be very generic and can 
be used to finite derivative gravity as well. Now expanding the action Eq.~(\ref{quad-action}) around Minkowski space up to terms 
containing $\mathcal{O}(h^2)$ contributions will help us to find the graviton propagator. 
\begin{eqnarray}\label{lin-act-0}
S_{q}= - \int d^4 x \left[\frac{1}{2}h_{\mu \nu} \Box a(\Box) h^{\mu
\nu}+h_{\mu}^{\sigma} b(\Box) \partial_{\sigma} \partial_{\nu} h^{\mu \nu} \right.
\nonumber \\
\left.+h c(\Box)\partial_{\mu} \partial_{\nu}h^{\mu \nu} + \frac{1}{2}h \Box d(\Box)h \right.\nonumber \\
\left.+ h^{\lambda \sigma} \frac{f(\Box)}{2\Box}\partial_{\sigma}\partial_{\lambda}\partial_{\mu}\partial_{\nu}h^{\mu
\nu}\right] \,.
\end{eqnarray}
The above equation and the form factors: $a(\Box)$, $b(\Box)$, $c(\Box)$, $d(\Box)$ and $f(\Box)$ 
are the same as first derived in Refs.~\cite{Biswas:2011ar,Biswas:2013kla}. 
All the contractions are due to  $\eta_{\mu \nu} \eta^{\mu \nu}$, and the expressions
for  $a(\Box)$, $b(\Box)$, $c(\Box)$, $d(\Box)$ and $f(\Box)$ will now read as:
\begin{align}
a(\Box)&=1-\frac{1}{2}\mathcal{F}_{2}(\Box)\frac{\Box}{M_s^2}-2\mathcal{F}_{3}(\Box)\frac{\Box}{M_s^2}, \\
b(\Box)&=-1+\frac{1}{2}\mathcal{F}_{2}(\Box)\frac{\Box}{M_s^2}+2\mathcal{F}_{3}(\Box)\frac{\Box}{M_s^2},\\
c(\Box)&=1+2\mathcal{F}_{1}(\Box)\frac{\Box}{M_s^2}+\frac{1}{2}\mathcal{F}_{2}(\Box)\frac{\Box}{M_s^2},\\
d(\Box)&=-1-2\mathcal{F}_{1}(\Box)\frac{\Box}{M_s^2}-\frac{1}{2}\mathcal{F}_{2}(\Box)\frac{\Box}{M_s^2},\\
f(\Box)&=-2\mathcal{F}_{1}(\Box)\frac{\Box}{M_s^2}-\mathcal{F}_{2}(\Box)\frac{\Box}{M_s^2}-2\mathcal{F}_{3}(\Box)\frac{\Box}{M_s^2}.
\end{align}
From the above expression, we can easily see that when we take $M_s\rightarrow \infty$, we reduces  to the local limit,
which in our case is pure GR, for which $a(\Box)=c(\Box)=1,~b(\Box)=d(\Box)=-1,~f(\Box)=0$ in the linearized action Eq.(\ref{lin-act-0}). From the above we note that 
\begin{align}
a(\Box)+b(\Box)&=0\,,
\\c(\Box)+d(\Box)&=0\,,
\\ b(\Box)+c(\Box)+f(\Box)&=0\,,
\end{align}
which is a consequence of Bianchi identity as shown in Refs.~\cite{Biswas:2011ar,Biswas:2013kla}. 
\begin{multline}\label{bianchi}
\nabla_{\mu}\tau^{\mu}_{\nu}=0=(c+d)\Box\partial_{\nu}h \\
 +(a+b)\Box h^{\mu}_{\nu,\mu}+(b+c+f)h^{\alpha\beta}_{,\alpha\beta\nu}\,,
\end{multline}
which verifies the constraints (A8-A10).
Without loss of generality we can assuming $f(\Box)=0$, then we obtain $a(\Box)=c(\Box)$, which we will show consistent 
with the expectations of GR propagator. This condition further
constraints the original form factor, $$2\mathcal{F}_{1}(\Box)+\mathcal{F}_{2}(\Box)+2\mathcal{F}_{3}(\Box)=0.$$

Now the spin projection operators for  tensor of rank $2$ can be analyzed in arbitrary $D$-dimensions. We can take the limit $D=4$ for the relevant case we are interested in here. Around the Minkowski spacetime we can write them as follows, see~\cite{Rivers:1964,VanNieuwenhuizen:1973fi,Biswas:2013kla}
\begin{equation}
\mathcal{P}^{2}=\frac{1}{2}(\theta_{\mu \rho}\theta_{\nu \sigma}+\theta_{\mu
\sigma}\theta_{\nu \rho} ) - \frac{1}{D-1}\theta_{\mu \nu}\theta_{\rho
\sigma},
\end{equation}
corresponds to spin-2 component. The vector component of the projection operator is given by:
\begin{equation}
\mathcal{P}^{1}=\frac{1}{2}( \theta_{\mu \rho}\omega_{\nu \sigma}+\theta_{\mu
\sigma}\omega_{\nu \rho}+\theta_{\nu \rho}\omega_{\mu \sigma}+\theta_{\nu
\sigma}\omega_{\mu \rho} ).
\end{equation}
There are 2 spin-0 operators
\begin{equation}
\mathcal{P}_{s}^{0}=\frac{1}{D-1}\theta_{\mu \nu} \theta_{\rho \sigma},
\end{equation}
\begin{equation}
\mathcal{P}_{w}^{0}=\omega_{\mu \nu}\omega_{\rho \sigma},
\end{equation}
and the mixing between the two spin-0 operators:
\begin{equation}
\mathcal{P}_{sw}^{0}=\frac{1}{\sqrt{D-1}}\theta_{\mu \nu}\omega_{\rho \sigma},
\end{equation}
\begin{equation}
\mathcal{P}_{ws}^{0}=\frac{1}{\sqrt{D-1}}\omega_{\mu \nu}\theta_{\rho \sigma},
\end{equation}
where
\begin{equation}
\theta_{\mu \nu}=\eta_{\mu \nu}-\frac{k_{\mu}k_{\nu}}{k^2}
\end{equation}
and
\begin{equation}
\omega_{\mu \nu}=\frac{k_{\mu}k_{\nu}}{k^2}.
\end{equation}
we  can show that 
\begin{equation}
a(\Box)h_{\mu \nu} \rightarrow\ a(-k^2)\left[\mathcal{P}^{2}+\mathcal{P}^{1}+\mathcal{P}_{s}^{0}+\mathcal{P}_{w}^{0}\right]h,
\end{equation}
\begin{equation}
b(\Box)\partial_{\sigma}\partial_{(\nu}h_{\mu)}^{\sigma} \rightarrow -b(-k^2)k^2\left[\mathcal{P}^1+2\mathcal{P}_{w}^{0}\right]h,
\end{equation}
\begin{multline}
c(\Box)(\eta_{\mu \nu}\partial_{\rho}\partial_{\sigma}h^{\rho \sigma}+\partial_{\mu}\partial_{\nu}h
 )\\ 
\rightarrow\  -c(-k^2)k^2 \left[2\mathcal{P}_{w}^{0}+\sqrt{D-1}\left(\mathcal{P}_{sw}^{0}+\mathcal{P}_{ws}^{0}\right)
\right]h,
\end{multline}
\begin{multline}
 \eta_{\mu \nu} d(\Box)h \\
\rightarrow\  d(-k^2)\left[(D-1)\mathcal{P}_{s}^{0}+\mathcal{P}_{w}^{0}+\sqrt{D-1}\left(\mathcal{P}_{sw}^{0}+\mathcal{P}_{ws}^{0}\right)
\right]h,
\end{multline}
\begin{equation}
f(\Box)\partial^{\sigma}\partial^{\rho}\partial_{\mu}\partial_{\nu}h_{\rho
\sigma} \rightarrow\ f(-k^2)k^4\mathcal{P}_{w}^{0}h.
\end{equation}
Hence,
\begin{equation}
ak^2\mathcal{P}^2h=\kappa \mathcal{P} ^2 \tau \Rightarrow\ \mathcal{P}^{2}h=\kappa\left(\frac{\mathcal{P}^2}{ak^2}\right)\tau,
\end{equation}
\begin{equation}
(a+b)k^2 \mathcal{P}^{1}h=\kappa\mathcal{P}^{1}\tau \Rightarrow\ \mathcal{P}^{1}\tau=0,
\end{equation}
\begin{equation}
(a+(D-1)d)k^2\mathcal{P}_{s}^{0}h+(c+d)k^2\sqrt{D-1}\mathcal{P}_{sw}^{0}h=\kappa\mathcal{P}_{s}^{0}\tau,
\end{equation}
\begin{equation}
(c+d)k^2\sqrt{D-1}\mathcal{P}_{ws}^{0}h+(a+2b+2c+d+f)k^2\mathcal{P}_{w}^{0}h=\kappa\mathcal{P}_{w}^{0}\tau.
\end{equation}
So, 
\begin{multline}
(a+(D-1)d)k^2\mathcal{P}_{s}^{0}h= \kappa \mathcal{P}_{s}^{0}\tau \\
\Rightarrow\ \mathcal{P}_{s}^{0}h=\kappa \frac{\mathcal{P}_{s}^{0}}{(a+(D-1)d)k^2}\tau,
\end{multline}
\begin{multline}
(a+2b+2c+d+f)k^{2}\mathcal{P}_{w}^{0}h= \kappa \mathcal{P}_{w}^{0} \tau\\
\Rightarrow\mathcal{P}_{w}^{0} h= \kappa \frac{\mathcal{P}_{w}^{0}}{(a+2b+2c+d+f)k^2}
\tau,
\end{multline}
where we have used the constraints given by (A10-A12). Note that the denominator corresponding to the $P_w^0$ spin projector vanishes so that there is no $w$-multiplet. The $D$-dimensional graviton propagator is given by
\begin{equation}
\label{fullprop}
\Pi(-k^{2})=\frac{\mathcal{P}^{2}}{k^{2}a(-k^{2})}+\frac{\mathcal{P}_{s}^{0}}{k^{2}(a(-k^{2})-(D-1)c(-k^{2}))}\,.
 \end{equation} 
 This is also known as the saturated or sandwiched propagator between two conserved currents, i.e. ${\cal J}_1\Pi{\cal J}_2$, where 
 ${\cal J}_i$ ($i=1,2$) are conserved currents, corresponding to the two vertices. Note that propagator always contains gauge dependent part, which is unphysical, and gauge independent part, which is physical. Namely the part of the propagator which contributes to the scattering amplitude, which is physical.
By assuming $f(\Box)=0\Rightarrow a(\Box)=c(\Box)$, which corresponds to no scalar propagating degree of freedom in the propagator, and the graviton propagator in any $D$ dimensions is given by~\cite{Biswas:2011ar,Biswas:2013kla}:
\begin{equation}
\Pi = \frac{1}{k^2a(-k^2)} ( \mathcal{P}^{2}-\frac{1}{D-2}\mathcal{P}_{s}^{0}).
\end{equation}
In $D=4$ dimensions, we obtain the standard graviton propagator for GR,  when we take $a(-k^2)=1$, since $a(\Box)=1$), we obtain the standard gauge independent part of the propagator which is physical~\cite{VanNieuwenhuizen:1973fi}: 
\begin{equation}
\Pi = \frac{1}{k^2} ( \mathcal{P}^{2}-\frac{1}{2}\mathcal{P}_{s}^{0}).
\end{equation}
The saturated graviton propagator is massless, and carries massless spin-0 and massless spin-2 components. Further note that for IDG we will have to keep $a(-k^2)$, such that it does not have any poles, means it has to be exponential of an entire function, which has no poles in the complex plane, i.e. $e^{\gamma(k^2)}$, where 
$\gamma(k^2)$ is an entire function, and has no dynamical degrees of freedom. Therefore, the IDG propagator has no new dynamical degrees of freedom other than that of GR, see for details~\cite{Biswas:2011ar}.

\section{Entropy Calculations \label{app:entropy}}

To calculate the entropy we begin by considering the entangled state
\begin{equation}
	\left|\psi\right\rangle=\frac{1}{2}\left(\left|\downarrow\downarrow\right\rangle+e^{i\Delta\theta}\left|\downarrow\uparrow\right\rangle+e^{i\Delta\phi}\left|\uparrow\downarrow\right\rangle+\left|\uparrow\uparrow\right\rangle\right)
\end{equation}

where $\Delta\theta=\frac{mG}{\hbar}\left(\Phi\left(r_{0}-\Delta x\right)-\Phi\left(r_{0}\right)\right)$ and $\Delta\phi=\frac{mG}{\hbar}\left(\Phi\left(r_{0}+\Delta x\right)-\Phi\left(r_{0}\right)\right)$ for an average distance between the interferometer arms $r_0$ and superposition size $\Delta x$. 

The density matrix defined as
\begin{equation}
	\hat{\rho}=\left|\psi\right\rangle\left\langle\psi\right|
\end{equation}

and the reduced density matrix is then
\begin{align}
	\hat{\rho}_{A}=&Tr_{B}\left[\hat{\rho}\right] \nonumber\\
	=&  \frac{1}{4}\left[ {\begin{array}{cc}
		2 & e^{i\Delta\theta}+e^{-i\Delta\phi} \\
		e^{-i\Delta\theta}+e^{i\Delta\phi} & 2 \\
		\end{array} } \right].
\end{align}

Now the eigenvalues of $\hat{\rho}_{A}$ can be used  to calculate the entropy, using the standard eigenvalue equation we have

\begin{align}
	0=&\left|\hat{\rho}_{A}-\lambda \mathbb{I}\right|\\
	0=&\left(\frac{1}{2}-\lambda\right)^{2}-\frac{1}{4}\left(e^{i\Delta\theta}+e^{-i\Delta\phi}\right)\frac{1}{4}\left(e^{-i\Delta\theta}+e^{i\Delta\phi}\right) \\
	\implies\lambda_{\pm}=&\frac{1}{2}\pm\frac{1}{2}\bigg[\frac{1}{2}\bigg(1+\cos\bigg(\frac{m\tau}{\hbar}(\Phi\left(r_0-\Delta x\right)\nonumber\\
	&+\Phi\left(r_0+\Delta x\right)-2\Phi\left(r_0\right))\bigg)\bigg)\bigg]^{1/2}
\end{align}

and so finally giving the entropy as

\begin{equation}
\mathcal{S}\left(\hat{\rho}_A\right)=-\left(\lambda_{-}\log_{2}\left(\lambda_{-}\right)+\lambda_{+}\log_{2}\left(\lambda_{+}\right)\right).
\end{equation}
\end{document}